# Structural and electronic properties of uranium-encapsulated $Au_{14}$ cage

Yang Gao, [1] Xing Dai, [1] Seung-gu Kang, [2] Camilo Andres Jimenez Cruz, [2] Minsi Xin, [1] Yan Meng, [1] Jie Han, [1] Zhigang Wang, [1,*] and Ruhong Zhou, [1,2,3,*]

[1] *Institute of Atomic and Molecular Physics, Jilin University, Changchun 130012, P. R. China*

[2] *Computational Biology Center, IBM Thomas J. Watson Research Center, Yorktown Heights, NY 10598*

[3] *Department of Chemistry, Columbia University, New York, NY 10027*

*To whom correspondence should be addressed. E-mail: wangzg@jlu.edu.cn; ruhongz@us.ibm.com

**Abstract:** The structural properties of the uranium-encapsulated nano-cage U@$Au_{14}$ are predicted using density functional theory. The presence of the uranium atom makes the $Au_{14}$ structure more stable than the empty $Au_{14}$-cage, with a triplet ground electronic state for U@$Au_{14}$. Analysis of the electronic structure shows that the two frontier single-occupied molecular orbital electrons of U@$Au_{14}$ mainly originate from the 5f shell of the U atom after charge transfer. Meanwhile, the bonding orbitals have both the 5f and 6d components of the U atom, along with the 5d and 6s components of the Au atoms, indicating the covalent nature of the interaction between the U and Au atoms. Moreover, the charge population analysis shows that this nanostructure displays some unique electronic properties where the encapsulated atom gains electrons while the outer shell loses electrons. Therefore, this designed U@$Au_{14}$ nano-cage structure is stabilized by ionocovalent interactions. The current findings provide theoretical basis for future syntheses and further study of actinide doped gold nanoclusters, which might subsequently facilitate applications of such structure in radio-labeling, nanodrug carrier and other biomedical applications.

Due to their capability of containing other molecules in addition to their large surface areas, hollow cage structures have gained substantial attention recently in the studies of metallic clusters [1]. These bimetallic nano-cages can subsequently form larger structures, which may potentially be applied to construct more stable core@shell nanoclusters [2]. Doped gold clusters have also become a hot subject of research because of their stable chemical activity and unique configuration [3-8]. The interactions between the encapsulated atoms and gold nano-cages allow fine tuning the formation of nano-particle agglomerates, resulting in more compact structures with lower energy [9-11], and thus making these encapsulated gold clusters more stable than the hollow ones. Moreover, the synergistic effect between bimetallic nanoparticles, and the great mutability of their components, structures and properties, could enhance some specific performance of metals, granting bimetallic gold clusters broad applications in fields such as bioengineering and biocatalysts [8].

Actinide elements are known to have active chemical properties and distinctive electronic structures which can be used to develop novel nanomaterials. Previous studies on actinide-encapsulated fullerenes [12-16] have not only helped further understand the complex electronic structures and chemical activities of the confined actinides, but have also stimulated the applications of actinides in functional nanomaterials and nanomedicine. Compared with fullerenes, gold nanoclusters have stronger resistance to oxidation, greater biocompatibility, higher density, and better photoelectric properties. Thus, gold-based bimetallic nano-clusters have emerged as an important line of research [8,17]. Besides adding an extra degree of freedom in the stoichiometry, doping gold clusters may be a powerful way to tune their chemical and physical properties [10,18]. Bioanalytical methods on such nanoparticles have also made substantial achievements in recent years [19-21].

In fact, to enhance the stability of gold clusters and improve their chemical activity, many theoretical and experimental studies have been done on encapsulating foreign atoms in the gold nano-cages.[10,22] Pyykko et al. first predicted the existence of the icosahedral W@$Au_{12}$ cluster. Since $Au_{12}$-cage itself is unstable [8], their study accredited the stability of this cluster to the aurophilic attraction, 18-electron rule and relativistic effects [3]. Quickly after this prediction, Li Xi et al. validated it by successfully synthesizing icosahedral W@$Au_{12}$ and Mo@$Au_{12}$ clusters under gaseous phase condition [4]. Analogously, the number of peripheral gold atoms has great influence on the location of the doped atom and cluster structure. Zeng et al. found that, in theory, gold clusters ($Au_n$) doped with a foreign metal atom tend to form core@shell structures when the number of gold atoms, n, is 9 or greater [23]. A series of endohedral gold-cage clusters were then successfully synthesized, including M@$Au_{16}^-$ (M=Cu, Fe, Co, Ni) and Cu@$Au_{17}^-$ by anion photoelectron spectroscopy [24] and $Au_{24}Pd_1$ cluster by employing aberration-corrected scanning transmission electron microscopy [25]. Along with these developments of small-sized gold clusters, medium-sized metal atom-encapsulated gold-cage structures, for example Zn@$Au_{20}$ and X@$Au_{32}$ (X=Li+, Na+, K+, Rb+, Cs+) [26] were also developed. Interestingly, it was recently predicted that when n=14, the gold-covered bimetallic cluster can achieve the highest binding energy per atom [23]. As previously reported [6], a highly stable bimetallic $Au_{14}$-cage cluster can exist, and this structure has a larger HOMO-LUMO gap than those of W@$Au_{12}$ and $Au_{20}$ [4,27], indicating that metal atom-encapsulated $Au_{14}$ possesses a unique electronic structure, which facilitates the higher stability.

Gold nanoclusters have wide applications in biomedical fields in recent years due to their high biocompatibility and low toxicity, and with encapsulated metal atoms, they can not only exhibit more stable structures (than hollow clusters), but also modify their own chemical and physical properties by encapsulating foreign atoms. This is of particular interest to the applications of actinide elements in nuclear medicine with much reduced potential cytotoxicity. This paper presents the study of the structural and electronic properties of gold clusters doped with the actinide U-atom based on the $Au_{14}$-cage structure. In order to thoroughly explore the influences of actinides encapsulation onto the properties of the gold-cage, uranium atoms containing unsaturated 6d and 5f shells were both included in our study for the encapsulation with $Au_{14}$-cage. Through first-principle density functional theory (DFT), detailed analyses are performed for the electron density, molecular orbitals (MOs), charge transfer and density of states (DOS) of the nanostructure. We hope this study will provide a theoretical basis for future syntheses and further study of actinide doped gold nanoclusters.

For systems containing actinides or other heavy elements, substantial electron correlations and intense relativistic effects have to be included, which are caused by the high-speed movements of electrons. With DFT, electron correlations can be handled effectively without significantly increasing the computational complexity, as demonstrated in its wide applications in actinides and related systems [13, 28]. Similarly, for gold clusters, the DFT method seems to work very well [29]. Therefore, our first-principle calculations were based on the DFT method using the spin-polarized generalized gradient approximation (GGA) with the Perdew-Burke-Ernzerhof (PBE) exchange-correlation functional. We also used the BP86 functional for further comparative validation. Geometry optimizations were performed at the scalar-relativistic zero-order regular approximations (ZORA) level, followed by single-point energy calculations with inclusion of the spin-orbital coupling effects via the spin-orbital ZORA approach. ZORA is an effective approximation method that obtains positive energy state two-component Hamiltonian through four-component Hamiltonian [30]. and it is now considered one of the most accurate theoretical methods to approximately handle relativistic effects. Atomic partial charges were calculated similar to previous reports using the Mulliken population analysis [13]. Single-electron wave functions were expanded using the TZ2P (a triple-ζ basis set with two sets of polarization functions) uncontracted Slater-type orbital (STO) basis set for all atoms [31]. Since chemical changes are mainly the effects of valence electrons, inner electrons were frozen in subsequent calculations, for example, Au atoms were frozen to the 4d shell (i.e., treating the $4f^{14}5s^25p^65d^{10}6s^1$ as valence electrons), and the U atoms to the 4f shell (i.e. treating the $5s^25p^65d^{10}5f^36s^26p^66d^17s^2$ as valence electrons). In molecular computations, the inner-shell electrons were calculated by the Dirac equation with no relaxation. All calculations are spin unrestricted, and geometric optimizations have been performed without imposing any symmetric constraints. All obtained structures were then analyzed with vibration frequency calculations at the same level to avoid imaginary frequencies and ensure the reliability of the results. All the calculations were performed using the ADF2012 Package [32].

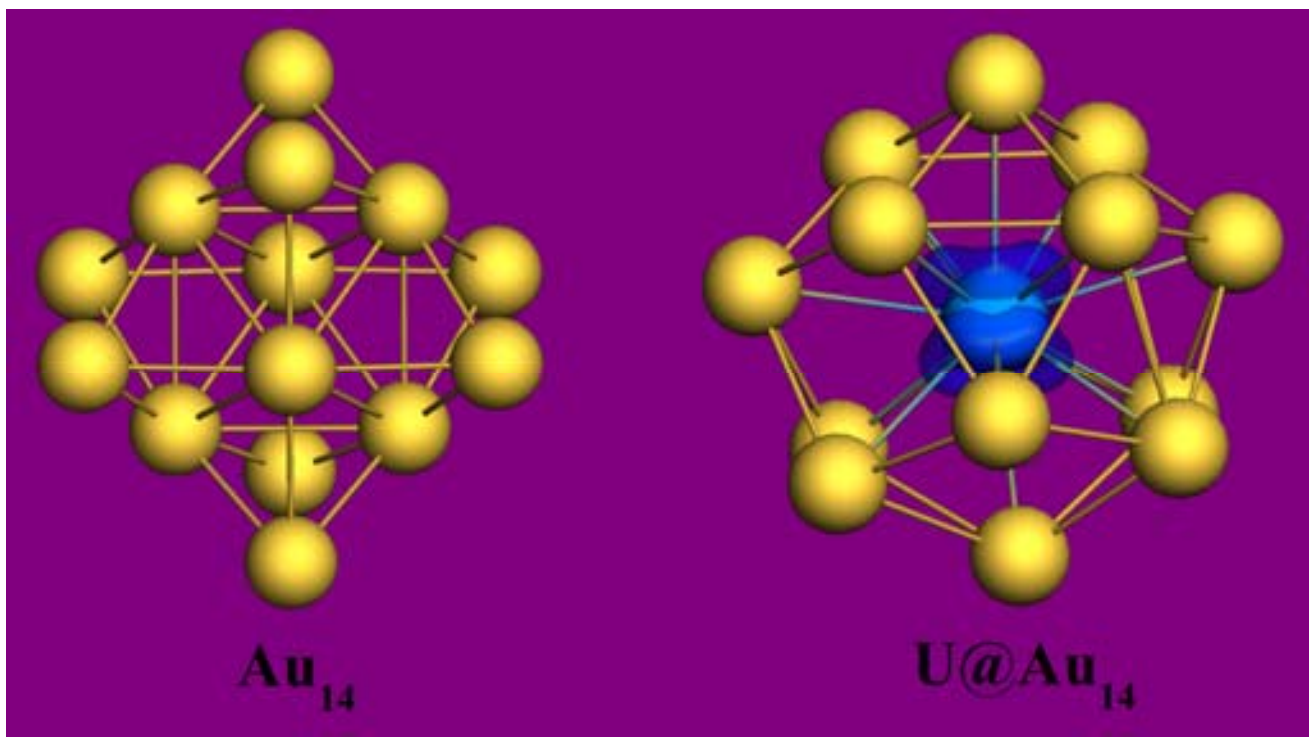

Figure 1. Stabilized structures of $Au_{14}$ and U@$Au_{14}$ after PBE/TZ2P level relaxation, and higher distribution of spin density (blue) exhibit on the U atom.

Table 1. Relative energies results (Scalar Relativistic) of U@ $Au_{14}$.

| System | Method[1] | Multiplicity | Bond energy (eV) | Bond energy + ZPE (eV) |
|---|---|---|---|---|
| U@$Au_{14}$ | PBE | 1 | 0.11 | 0.31 |
| | | 3 | 0.00 | 0.22 |
| | | 5 | 1.39 | 1.59 |
| | | 7 | 2.59 | 2.77 |
| | BP86 | 1 | 0.05 | 0.25 |
| | | 3 | 0.00 | 0.21 |
| | | 5 | 1.26 | 1.46 |
| | | 7 | 2.41 | 2.59 |

[1] The data above are calculated using PBE and BP86 functionals, respectively. The others in main body are all from PBE functional, unless specified otherwise.

Stable cage structures of $Au_{14}$ and U@$Au_{14}$ were obtained through PBE optimization, as shown in Figure 1, and further validated with the BP86 functional. The different spin multiplicities of U@$Au_{14}$ were calculated, and the ground state, determined by the lowest bond energy, is triplet (see Table 1). Energies of revised spin-orbital coupling effects were calculated and can be found in the Supporting Information (SI) Part 1. Next, the results were consistent for the two exchange-correlation functionals, PBE and BP86, with the triplet state displaying the lowest energy. After zero-point energy (ZPE) correction, the ground electronic state remained unchanged.

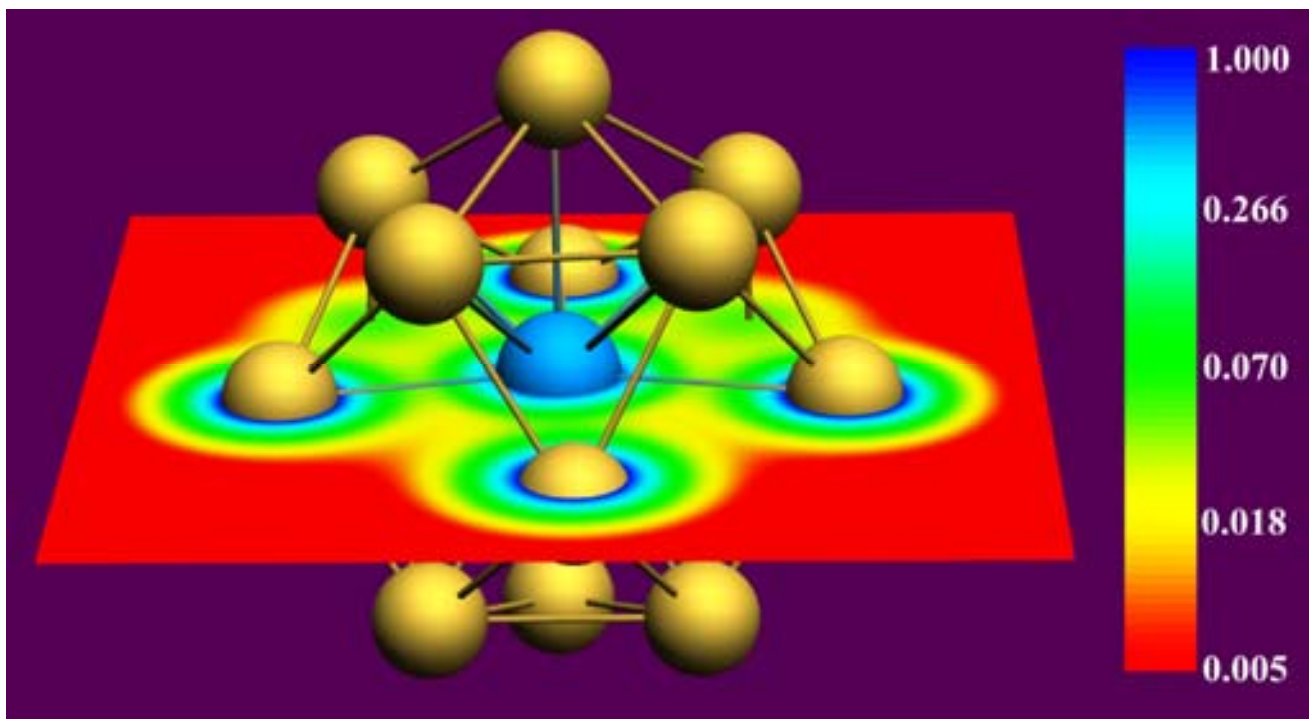

Figure 2. Color-filled map of electron density of U@$Au_{14}$.

The average bond energy per atom for pure $Au_{14}$-cage and U@$Au_{14}$ were 2.27 eV and 2.98 eV, respectively. Since structures with bond energies greater than 2.40 eV are commonly considered stable [23], our results suggest that U@$Au_{14}$ might have a very stable structure similar to those previously reported [1,9,33]. Another critical factor representing the stability of the U@$Au_{14}$ structure is the charge transfer between the internal uranium atom and peripheral gold atoms. The Mulliken charge distribution of the doped cage (SI Part 2) shows that electrons transfer from the peripheral Au atoms to the U atom, revealing the ionic character of the U-Au bond. The charge transferred from the Au atoms to the U atom originates primarily from the 6s and 5d electrons. As an illustration, the Mulliken populations in U@$Au_{14}$ were found to be $6s^{0.91}5d^{9.58}$ on Au and $7s^{0.53}5f^{2.97}6d^{3.71}$ on U (approximate electron occupation). To further examine the accuracy of this Mulliken method for describing the charge transfer in the U@$Au_{14}$ system, we performed additional calculations for cage structures with different numbers of surrounding Au atoms on the U atom (SI Part 3) and found that the number of Au atoms determines the direction of charge transfer, with the crossover at ~8 Au atoms. When the number of Au atoms is less than 8, the U atom is positively charged (i.e., electrons transferred from U to Au), but once it is larger than 8, the U atom is negatively charged (i.e., electrons transferred from Au to U). This pattern is consistent with those found in previous studies that the charge transfers from La atom to Au atoms at n=2-4, but the charge transfer direction is reversed at n=5 [34].

It is well known that the ground state of the isolated U atom is a quintet spin state, and the spin parallel electrons are all from the 5f and 6d shells of U [35]. However, the spin ground state of the system became triplet when the U atom is confined in the $Au_{14}$-cage, indicating that Au-confinement induces spin polarization. The inherent intense relativistic effects of the noble metal gold cause apparent hybridization between the inner-shell d orbitals and the outer-shell s orbitals (this hybrid orbitals are named 5d-6s orbitals). When foreign atoms are doped, hybridization becomes more apparent between 5d-6s orbitals and the valence electron orbitals of the doped atoms [3, 36]. A color-filled map of the electron density for the U@$Au_{14}$ structure, Figure 2, displays obvious electron accumulation (green areas) between the peripheral Au atoms and the U atom in the center, indicating covalent interaction. Subsequent orbital analysis demonstrates the hybridization between states directly. It should be noted that, although two sets of MOs (α and β orbitals) were calculated using the spin unrestricted method in this study, very few differences were found between them in their energy and shape aspects. Thus, the following double-occupied MOs of the system were represented by α orbitals only. Detailed contributions of the two sets MOs (α and β orbitals) are presented in SI Part 4.

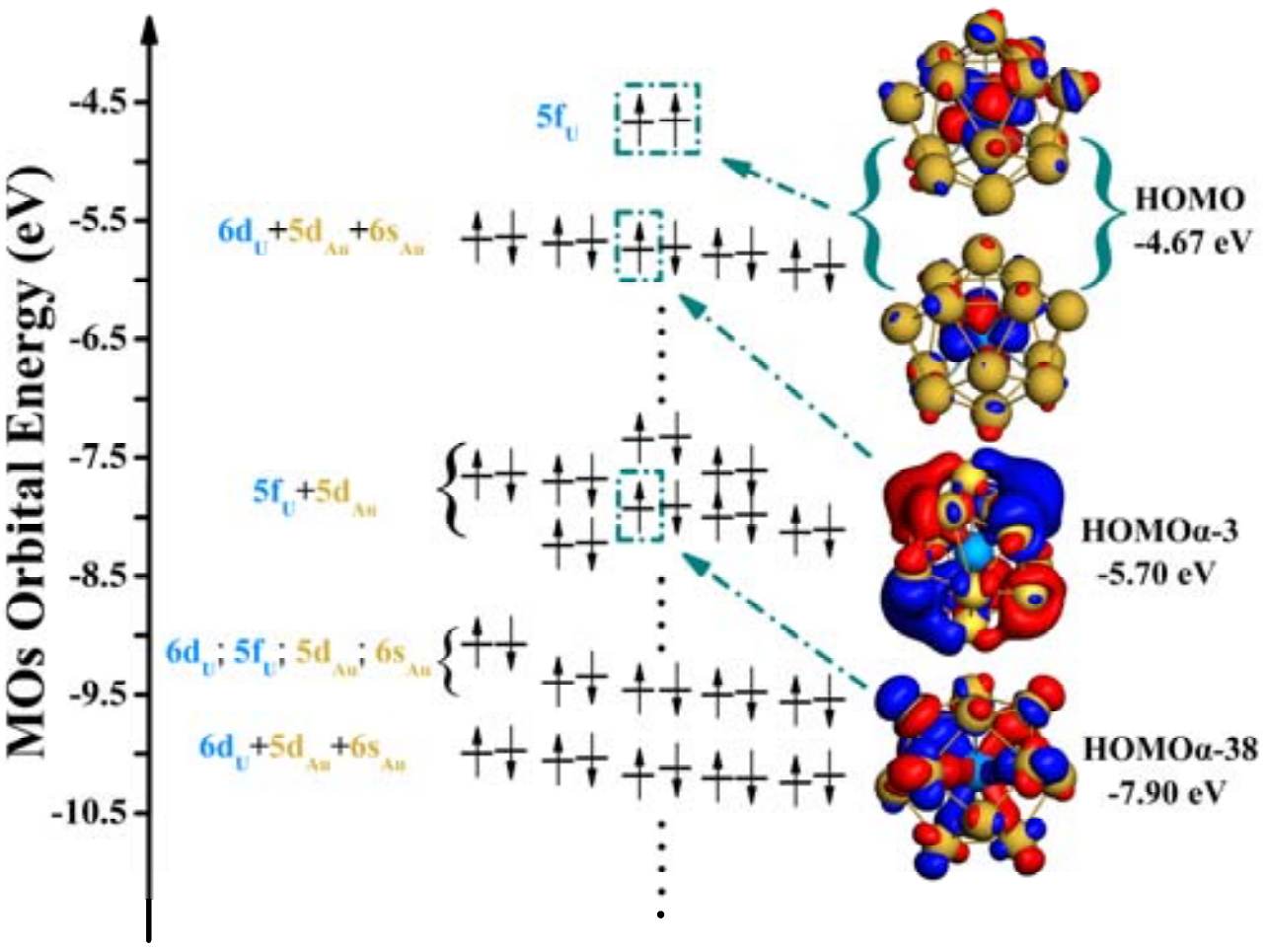

Figure 3. Electronic energy level diagram of ground state U@$Au_{14}$. The occupied MOs contributed from the 5f and 6d electrons of the U atom are listed in the diagram with typical MOs presented on the right. (Other occupied orbitals are presented in SI Part 5, isodensity = 0.02 au)

Table 2. Percentages of the 5f and 6d atomic orbitals of the U atom that are active in the intermolecular interaction.

| U@$Au_{14}$ (Triplet) | Uranium 5f | $Au_{14}$ 6d | Uranium 5f | $Au_{14}$ 6d |
|---|---|---|---|---|
| HOMO | 84.69% | 0% | 0% | 0% |
| HOMO | 89.33% | 0% | 0% | 0% |
| HOMOα-1 | 0%[1] | 9.87% | 32.76% | 17.17% |
| HOMOα-2 | 0% | 10.64% | 30.15% | 16.70% |
| HOMOα-3 | 0% | 9.36% | 32.42% | 12.52% |
| HOMOα-4 | 0% | 10.52% | 32.86% | 10.41% |
| HOMOα-5 | 0% | 9.24% | 28.71% | 18.96% |
| HOMOβ-1 | 0% | 9.17% | 33.13% | 16.69% |
| HOMOβ-2 | 0% | 9.26% | 30.70% | 15.64% |
| HOMOβ-3 | 0% | 8.91% | 32.39% | 14.27% |
| HOMOβ-4 | 0% | 9.85% | 32.87% | 11.45% |
| HOMOβ-5 | 0% | 8.73% | 29.52% | 18.56% |

[1] Percentages smaller than 1% are shown as 0% in the table.

Since 5f and 6d atomic orbitals of the isolated U atom are not fully occupied by electrons, they contribute significantly to bonding [35]. Therefore, in this study we focus on the contributions of these orbitals. The spin triplet suggests that two unpaired electrons exist in the system, and it can be seen from Figure 3 that the U atom has significant contributions in single-occupied MOs. Direct contributions from atomic orbitals to MOs, listed in Table 2, show that the 5f shell electrons account for 85.61% and 86.90% of the contributions to the two single-electron-occupied orbitals (classified as degenerate orbitals). On the other hand, for the two-electron occupied orbitals HOMO-1 to HOMO-5, their shapes reflect that of "covalent bonds" between the U atom and Au-cage. For example, the HOMOα-3 orbital is composed of the 6d components of U and the 6s and 5d components of Au. According to SI Part 4, there are several other molecular orbitals that are composed of the 5f components of U and 5d components of Au. The shape of HOMOα-38 orbital also suggests that covalent interaction exists between the U and Au atoms. Moreover, from HOMO-58 to HOMO-78 MOs, in addition to the hybridization forms described above, some orbitals consist of 5f, 6d, 5d and 6s components in aggregate (See SI Part 5, HOMO-67 and HOMO-68). Therefore, the two single-occupied electrons of the system originate mainly from the 5f shell of the U atom, while in the double-occupied orbitals, the electrons of the 5f and 6d shell of U all participate in U-Au bonding. From our present analysis, we conclude that the $U@Au_{14}$ structure is stabilized by ionocovalent interactions.

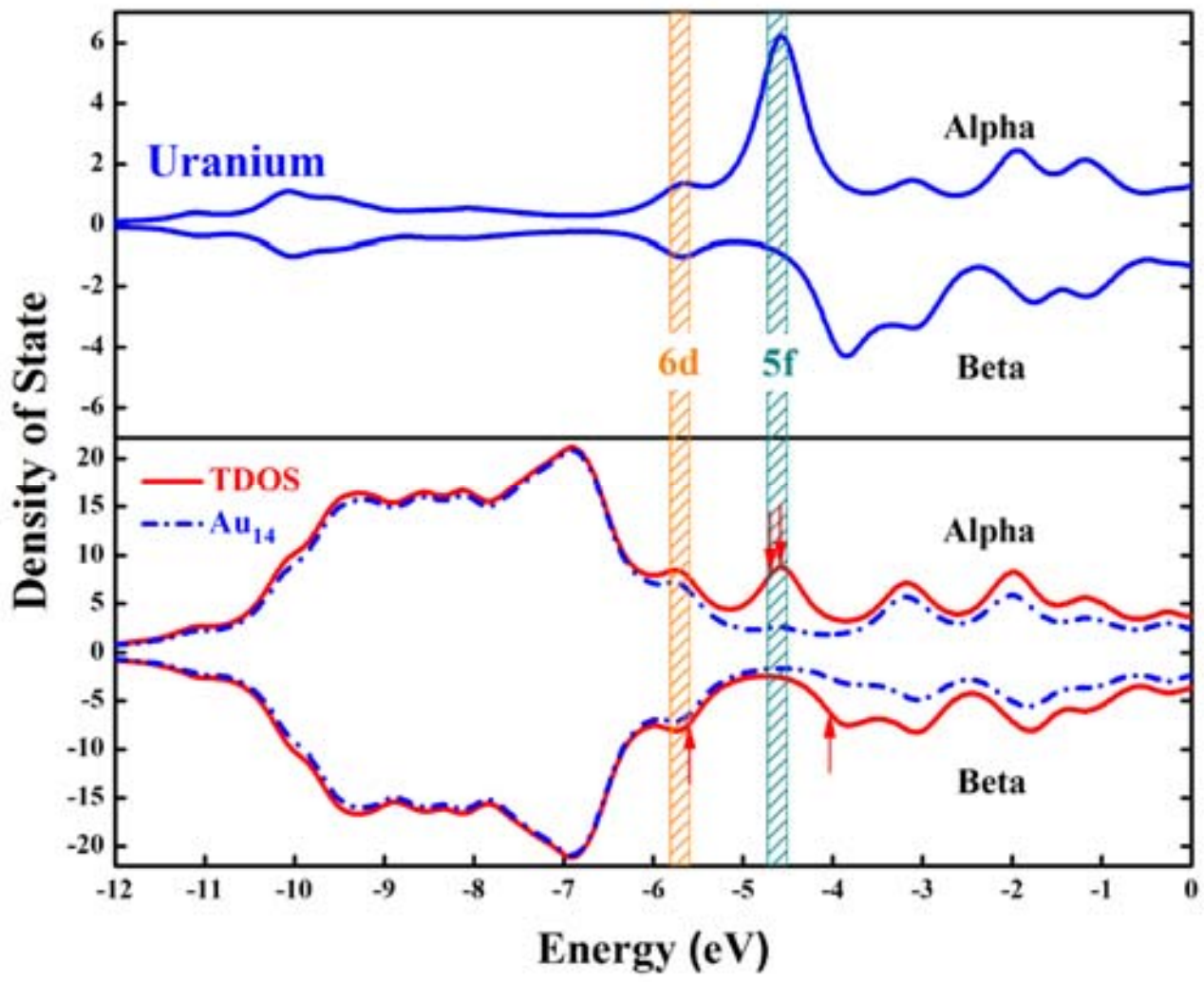

Figure 4. Density of states of $U@Au_{14}$. Arrows indicate the locations of HOMO/LUMO. The yellow brown shadow and the green shadow represent the 6d and 5f (the occupied and non-occupied states near the frontier orbitals) components of U respectively.

To further understand the bonding mechanism between the U atom and gold-cage, the DOS of the system was calculated (Figure 4). First, it is apparent that the majority positions of the two lines (TDOS: total density of state, red full line; LDOS: local density of state, blue dotted line) are concurrent between -12.00 eV and -6.00 eV, indicating that the contribution in this part is mainly from the Au-cage. The TDOS exhibits two sharp local peaks between -6.00 eV and -4.50 eV. The first peak is around -5.70 eV, the second one around -4.70 eV. Comparing TDOS with the LDOS of the U atom, the first peak is contributed by the U atom and Au-cage jointly, which indicates the intense electron interaction/sharing between them. On the other hand, the second peak is mainly contributed by the U atom. In addition, the orbital component analysis above showed that the two peaks correspond to the MOs from HOMO to HOMO-5. These results also indicate that the electrons of 6d and 5f shells participate in the intense interaction between the U atom and gold-cage.

Origins of the electronic states of the system are presented clearly in the spin density map (Figure 1), higher spin density distribution exists on the U atom. Analysis shows that the spin contribution of the U atom is spin=2.03. The reason why the ground state of the $U@Au_{14}$ structure is triplet is that the two unpaired electrons on the U atoms exist in the form of parallel spin. Through these analyses, we can confirm that the $U@Au_{14}$ structure is different from ordinary systems of which the inner metal and outer shell participate jointly in the electronic state adjustment, such as $U_2@C_{61}$ [13], instead it is a self-correcting system after electron transfer between them within the confinement. Meanwhile, the $U@Au_{14}$ nanostructure displays some unique electronic properties with the encapsulated metal atom gaining electrons and the outer cluster losing them, while the majority of such nanostructures show the opposite electron transfers [13-15].

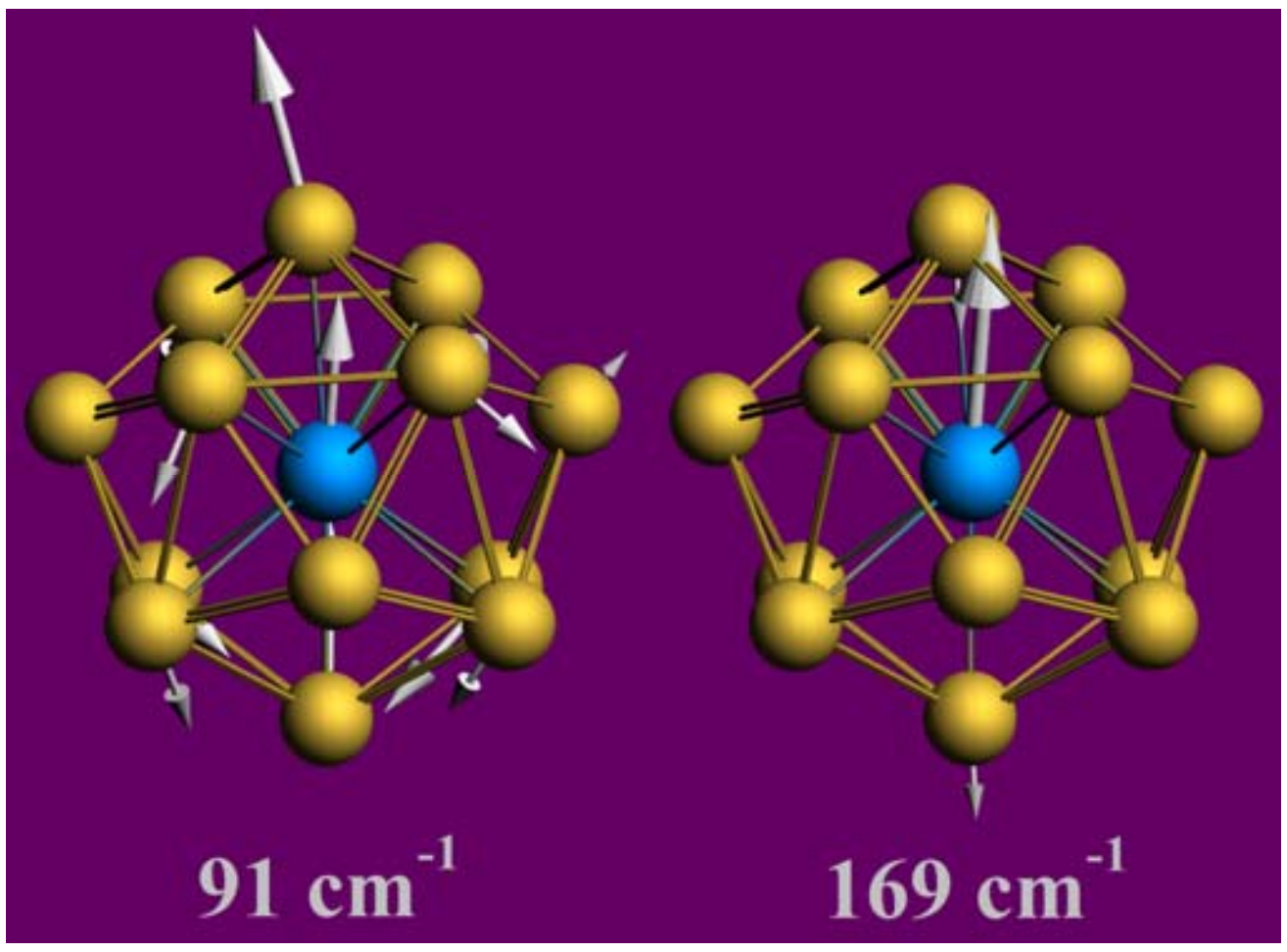

Figure 5. Two typical vibration modes of $U@Au_{14}$.

Finally, the vibrational spectrum contains two typical vibration modes with frequencies of 91 cm$^{-1}$ and 169 cm$^{-1}$, corresponding to the collaborative vibration between U and Au, which are presented in Figure 5. The infrared spectra of different clusters are determined by vibration modes of different atoms, and reflect the inherent properties of clusters. Thus, the vibration modes of $U@Au_{14}$ structure also indicate the covalent bonding nature between the Au-cage and U atom inside.

To summarize, we have designed a novel uranium-embedded nano-cage structure $U@Au_{14}$ based on a typical $Au_{14}$ cluster and the corresponding analysis is presented here. Because of the distinctive bonding characters of the 5f and 6d electrons of actinides elements, they exhibit interesting electronic properties along with their compounds. With the first-principle DFT, the ground electronic state of $U@Au_{14}$ is found to be a triplet. Orbital analysis showed that the two frontier single-electron orbitals are mainly contributed by the 5f electron of U atom, indicating that the $Au_{14}$-cage restrains the chemical toxicity of U atom. These single-electron MOs also maintain its original atomic orbital characters, and exhibit similar characteristics to those of the endohedral metallofullerenes (EMFs) or passivated Quantum Dots (QDs) [37]. In addition, the bonding orbitals of the gold-cage and the U atom have components of both 5f and 6d. More specifically, intense electronic interactions between U and Au are observed in the DOS, with charge transfer between the inner metal and outer shell, as well as the presence of collaborative vibration modes.

The $U@Au_{14}$ structure is a self-correcting system after electron transfer between them within the confinement. This feature might be useful in biomedicine, where gold nanoclusters are often used in therapeutics [20]. By encapsulating foreign metal atoms, the structural stability of gold nanoclusters can be further enhanced, and their electronic structure can also be adjusted. It means that their magnetism can be fine-tuned, facilitating wide applications in bioseparation, drug receptor targeting, and tumor hyperthermia [8]. In recent years, Gd-encapsulated gold-cage structure has been constructed experimentally, and successfully applied to Magnetic Resonance Imaging (MRI) [38]. Furthermore, the existence Gd-encapsulated gold-cage structure containing the 4f electron has been predicted theoretically [39]. However, it is remains unclear about the electron properties of actinides-encapsulated gold-cage structure containing the 5f electron. Through the current comprehensive first-principle studies on U atom-encapsulated gold-cage structure, we hope to provide a theoretical basis for future syntheses of such U-embedded nano-cage structures.

We wish to thank the support of the National Science Foundation of China under grant nos. 11374004 and 11004076. Z. W. also acknowledges the assistance of the High Performance Computing Center (HPCC) of Jilin University.

# Supporting Information

Part 1. Energy calculation for the system using spin-orbital ZORA approach.
Part 2. Mulliken populations of U@$Au_{14}$
Part 3. The charge transfer of system
Part 4. List of portion MOs, ordered by energy, with the most significant SFO gross populations
Part 5. Occupied MOs contributed by the 5f and 6d electrons of the U atom (α orbitals only)

**Part 1. Energy calculation to the system using spin-orbital ZORA approach.**

| System | Method | Multiplicity | Bond energy (eV) |
|---|---|---|---|
| U@$Au_{14}$ | PBE | **3** | **-101.80** |
| | | 5 | -101.36 |
| | | 7 | -101.27 |

On the basis of scalar-relativistic ZORA approach optimization system, energy calculation is performed for the corresponding structures using the spin-orbital coupling ZORA approach.

**Part 2. Mulliken populations of U@$Au_{14}$.**

| Number/Atom | Charge (e) | Spin density | Spin direction | s | p | d | f |
|---|---|---|---|---|---|---|---|
| 1 Au | 0.12 | 0.01 | α | 1.45 | 3.19 | 4.79 | 7.01 |
| | | | β | 1.45 | 3.18 | 4.79 | 7.01 |
| 2 Au | 0.12 | -0.00 | α | 1.46 | 3.18 | 4.79 | 7.01 |
| | | | β | 1.46 | 3.17 | 4.79 | 7.01 |
| 3 Au | 0.13 | -0.01 | α | 1.45 | 3.18 | 4.78 | 7.01 |
| | | | β | 1.46 | 3.18 | 4.79 | 7.01 |
| 4 Au | 0.12 | -0.00 | α | 1.46 | 3.18 | 4.79 | 7.01 |
| | | | β | 1.46 | 3.17 | 4.79 | 7.01 |
| 5 Au | 0.13 | -0.00 | α | 1.45 | 3.19 | 4.79 | 7.01 |
| | | | β | 1.45 | 3.19 | 4.79 | 7.01 |
| 6 Au | 0.13 | 0.01 | α | 1.45 | 3.18 | 4.79 | 7.01 |
| | | | β | 1.45 | 3.18 | 4.79 | 7.01 |
| 7 Au | 0.12 | -0.01 | α | 1.46 | 3.18 | 4.78 | 7.01 |
| | | | β | 1.46 | 3.18 | 4.79 | 7.01 |
| 8 Au | 0.11 | -0.01 | α | 1.45 | 3.20 | 4.78 | 7.01 |
| | | | β | 1.45 | 3.20 | 4.79 | 7.01 |
| 9 U | -1.68 | 2.03 | α | 2.27 | 6.24 | 6.90 | 2.45 |
| | | | β | 2.26 | 6.23 | 6.81 | 0.52 |
| 10 Au | 0.14 | -0.01 | α | 1.46 | 3.17 | 4.78 | 7.01 |
| | | | β | 1.46 | 3.17 | 4.79 | 7.01 |
| 11 Au | 0.12 | -0.00 | α | 1.45 | 3.18 | 4.80 | 7.01 |
| | | | β | 1.45 | 3.18 | 4.80 | 7.01 |
| 12 Au | 0.14 | -0.01 | α | 1.46 | 3.17 | 4.78 | 7.01 |
| | | | β | 1.46 | 3.17 | 4.79 | 7.01 |
| 13 Au | 0.08 | -0.01 | α | 1.46 | 3.18 | 4.79 | 7.01 |
| | | | β | 1.47 | 3.18 | 4.80 | 7.01 |
| 14 Au | 0.12 | 0.01 | α | 1.45 | 3.18 | 4.81 | 7.01 |
| | | | β | 1.45 | 3.17 | 4.80 | 7.01 |
| 15 Au | 0.10 | -0.00 | α | 1.46 | 3.18 | 4.80 | 7.01 |
| | | | β | 1.46 | 3.18 | 4.80 | 7.01 |

**Part 3. The charge transfer of system**

| **U@$Au_4$** | |
|---|---|
| Number/Atom | Charge (e) |
| 1 Au | -0.22 |
| 2 Au | -0.21 |
| 3 Au | -0.22 |
| 4 Au | -0.22 |
| **5 U** | **0.87** |
| **U@$Au_6$** | |
| Number/Atom | Charge (e) |
| 1 Au | -0.05 |
| 2 Au | -0.17 |
| 3 Au | -0.19 |
| 4 Au | -0.04 |
| 5 Au | -0.17 |
| 6 Au | -0.04 |
| **7 U** | **0.66** |
| **U@$Au_8$** | |
| Number/Atom | Charge (e) |
| 1 Au | 0.01 |
| 2 Au | 0.01 |
| 3 Au | 0.01 |
| 4 Au | 0.01 |
| 5 Au | 0.01 |
| 6 Au | 0.01 |
| 7 Au | 0.01 |
| 8 Au | 0.01 |
| **9 U** | **-0.08** |
| **U@$Au_{12}$** | |
| Number/Atom | Charge (e) |
| 1 Au | 0.10 |
| 2 Au | 0.11 |
| 3 Au | 0.11 |
| 4 Au | 0.10 |
| 5 Au | 0.10 |
| 6 Au | 0.11 |
| 7 Au | 0.10 |
| 8 Au | 0.10 |
| 9 Au | 0.10 |
| 10 Au | 0.10 |
| 11 Au | 0.10 |
| 12 Au | 0.10 |
| **13 U** | **-1.23** |
| **U@$Au_{16}$** | |
| 1 Au | 0.17 |
| 2 Au | 0.17 |
| 3 Au | 0.09 |
| 4 Au | 0.17 |
| 5 Au | 0.09 |
| 6 Au | 0.09 |
| 7 Au | 0.09 |
| 8 Au | 0.09 |
| 9 Au | 0.09 |
| 10 Au | 0.09 |
| 11 Au | 0.09 |
| 12 Au | 0.15 |
| 13 Au | 0.09 |
| 14 Au | 0.09 |
| 15 Au | 0.09 |
| 16 Au | 0.09 |
| **17 U** | **-1.74** |

All structures are relaxed to the status of lowest energy in PBE/TZ2P level. All obtained structures were then analyzed with vibration frequency calculations at the same level to avoid imaginary frequencies and ensure the reliability of the results. For the M@$Au_n$ structure, Ref S1 pointed out that it was more stable when the number of gold atoms n is even. Therefore, n was given the values of 4, 6, 8, 12 and 16 for comparison. Calculation shows that with the increment of n, the direction of charge transfer changes, and this pattern is consistent with those found in previous studies [34,S2].

**Part 4. List of portion MOs, ordered by energy, with the most significant SFO gross populations.**

| E (eV) | MO | % | SFO | Fragment |
|---|---|---|---|---|
| -5.63 | HOMOα-1 | 9.87% | **6d** | **U** |
| | | 32.76% | 6s | $Au_{14}$ |
| | | 17.17% | 5d | $Au_{14}$ |
| -5.67 | HOMOα-2 | 10.64% | **6d** | **U** |
| | | 30.15% | 6s | $Au_{14}$ |
| | | 16.70% | 5d | $Au_{14}$ |
| -5.70 | HOMOα-3 | 9.36% | **6d** | **U** |
| | | 32.42% | 6s | $Au_{14}$ |
| | | 12.52% | 5d | $Au_{14}$ |
| -5.73 | HOMOα-4 | 10.52% | **6d** | **U** |
| | | 32.86% | 6s | $Au_{14}$ |
| | | 10.41% | 5d | $Au_{14}$ |
| -5.83 | HOMOα-5 | 9.24% | **6d** | **U** |
| | | 28.71% | 6s | $Au_{14}$ |
| | | 18.96% | 5d | $Au_{14}$ |
| -7.36 | HOMOα-30 | 1.18% | **5f** | **U** |
| | | 85.48% | 5d | $Au_{14}$ |
| -7.60 | HOMOα-34 | 1.23% | **5f** | **U** |
| | | 81.76% | 5d | $Au_{14}$ |
| -7.63 | HOMOα-35 | 3.00% | **5f** | **U** |
| | | 79.36% | 5d | $Au_{14}$ |
| -7.68 | HOMOα-37 | 1.69% | **5f** | **U** |
| | | 81.30% | 5d | $Au_{14}$ |
| -7.89 | HOMOα-38 | 7.62% | **5f** | **U** |
| | | 77.95% | 5d | $Au_{14}$ |
| -7.98 | HOMOα-39 | 3.33% | **5f** | **U** |
| | | 81.18% | 5d | $Au_{14}$ |
| -8.09 | HOMOα-41 | 1.36% | **5f** | **U** |
| | | 79.79% | 5d | $Au_{14}$ |
| -8.19 | HOMOα-45 | 1.49% | **5f** | **U** |
| | | 80.45% | 5d | $Au_{14}$ |
| -9.06 | HOMOα-58 | 1.08% | **5f** | **U** |
| | | 81.76% | 5d | $Au_{14}$ |
| -9.40 | HOMOα-66 | 1.44% | **5f** | **U** |
| | | 6.86% | 6s | $Au_{14}$ |
| | | 69.18% | 5d | $Au_{14}$ |
| -9.50 | HOMOα-67 | 1.18% | **6d** | **U** |
| | | 1.24% | **5f** | **U** |
| | | 7.35% | 6s | $Au_{14}$ |
| | | 68.24% | 5d | $Au_{14}$ |
| -9.52 | HOMOα-68 | 1.49% | **6d** | **U** |
| | | 1.21% | **5f** | **U** |
| | | 6.17% | 6s | $Au_{14}$ |
| | | 74.37% | 5d | $Au_{14}$ |
| -9.53 | HOMOα-69 | 1.05% | **5f** | **U** |
| | | 1.19% | 6s | $Au_{14}$ |
| | | 77.12% | 5d | $Au_{14}$ |
| -9.97 | HOMOα-74 | 15.91% | **6d** | **U** |
| | | 7.21% | 6s | $Au_{14}$ |
| | | 56.87% | 5d | $Au_{14}$ |
| -10.05 | HOMOα-75 | 17.14% | **6d** | **U** |
| | | 9.54% | 6s | $Au_{14}$ |
| | | 53.01% | 5d | $Au_{14}$ |
| -10.13 | HOMOα-76 | 15.66% | **6d** | **U** |
| | | 10.89% | 6s | $Au_{14}$ |
| | | 58.06% | 5d | $Au_{14}$ |
| -10.14 | HOMOα-77 | 14.39% | **6d** | **U** |
| | | 11.04% | 6s | $Au_{14}$ |
| | | 52.04% | 5d | $Au_{14}$ |
| -10.16 | HOMOα-78 | 15.80% | **6d** | **U** |
| | | 12.32% | 6s | $Au_{14}$ |
| | | 47.32% | 5d | $Au_{14}$ |
| | | | | |
| -5.61 | HOMOβ-1 | 9.17% | **6d** | **U** |
| | | 33.13% | 6s | $Au_{14}$ |
| | | 16.69% | 5d | $Au_{14}$ |
| -5.65 | HOMOβ-2 | 9.26% | **6d** | **U** |
| | | 30.70% | 6s | $Au_{14}$ |
| | | 15.64% | 5d | $Au_{14}$ |
| -5.68 | HOMOβ-3 | 8.91% | **6d** | **U** |
| | | 32.39% | 6s | $Au_{14}$ |
| | | 14.27% | 5d | $Au_{14}$ |
| -5.71 | HOMOβ-4 | 9.85% | **6d** | **U** |
| | | 32.87% | 6s | $Au_{14}$ |
| | | 11.45% | 5d | $Au_{14}$ |
| -5.82 | HOMOβ-5 | 8.73% | **6d** | **U** |
| | | 29.52% | 6s | $Au_{14}$ |
| | | 18.56% | 5d | $Au_{14}$ |
| -7.60 | HOMOβ-34 | 1.10% | **5f** | **U** |
| | | 82.71% | 5d | $Au_{14}$ |
| -7.62 | HOMOβ-35 | 1.19% | **5f** | **U** |
| | | 81.22% | 5d | $Au_{14}$ |
| -7.66 | HOMOβ-37 | 1.31% | **5f** | **U** |
| | | 80.77% | 5d | $Au_{14}$ |
| -7.85 | HOMOβ-38 | 5.26% | **5f** | **U** |
| | | 81.46% | 5d | $Au_{14}$ |
| -7.96 | HOMOβ-39 | 2.69% | **5f** | **U** |
| | | 80.96% | 5d | $Au_{14}$ |
| -9.36 | HOMOβ-64 | 1.04% | **5f** | **U** |
| | | 2.33% | 6s | $Au_{14}$ |
| | | 73.55% | 5d | $Au_{14}$ |
| -9.47 | HOMOβ-67 | 1.45% | **6d** | **U** |
| | | 7.19% | 6s | $Au_{14}$ |
| | | 73.71% | 5d | $Au_{14}$ |
| -9.52 | HOMOβ-69 | 1.11% | **6d** | **U** |
| | | 1.60% | 6s | $Au_{14}$ |
| | | 77.20% | 5d | $Au_{14}$ |
| -9.34 | HOMOβ-74 | 14.67% | **6d** | **U** |
| | | 8.59% | 6s | $Au_{14}$ |
| | | 56.48% | 5d | $Au_{14}$ |
| -10.00 | HOMOβ-75 | 14.84% | **6d** | **U** |
| | | 8.30% | 6s | $Au_{14}$ |
| | | 56.18% | 5d | $Au_{14}$ |
| -10.07 | HOMOβ-76 | 13.47% | **6d** | **U** |
| | | 12.01% | 6s | $Au_{14}$ |
| | | 59.11% | 5d | $Au_{14}$ |
| -10.11 | HOMOβ-77 | 13.88% | **6d** | **U** |
| | | 10.12% | 6s | $Au_{14}$ |
| | | 51.13% | 5d | $Au_{14}$ |
| -10.13 | HOMOβ-78 | 14.42% | **6d** | **U** |
| | | 11.11% | 6s | $Au_{14}$ |
| | | 49.99% | 5d | $Au_{14}$ |

**Part 5. Occupied MOs contributed by the 5f and 6d electrons of the U atom (α orbitals only)**

| | | |
|---|---|---|
| 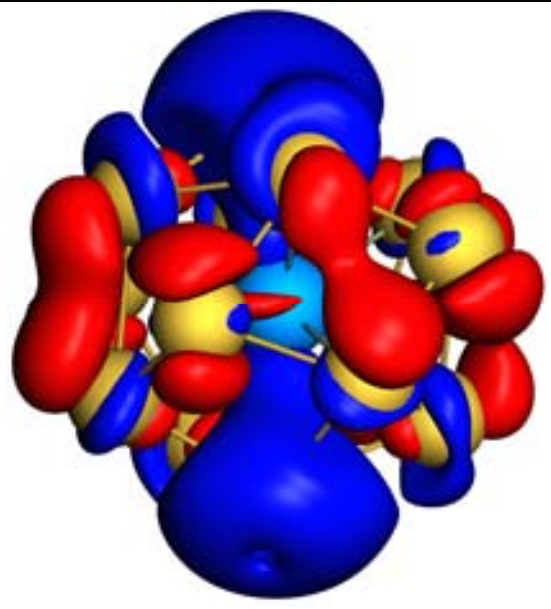<br>HOMOα-1 | 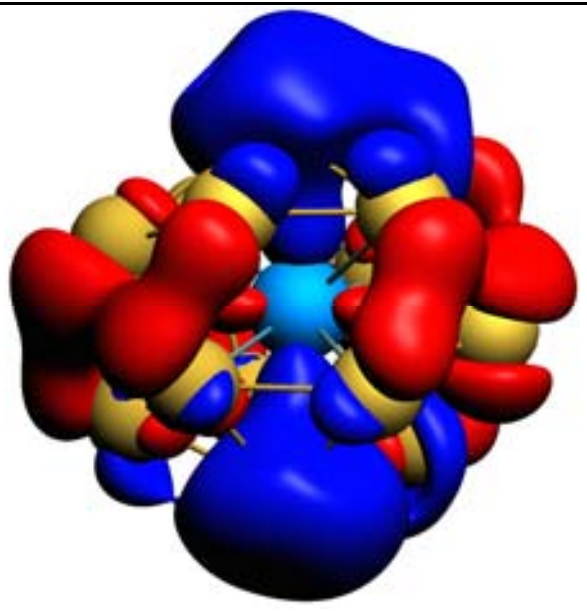<br>HOMOα-2 | 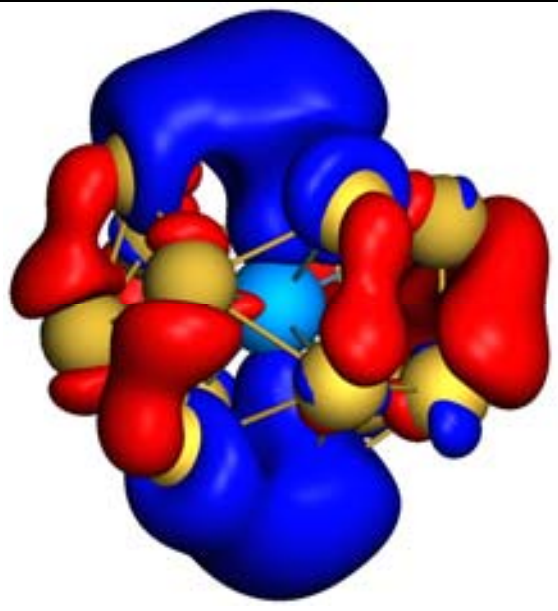<br>HOMOα-4 |
| 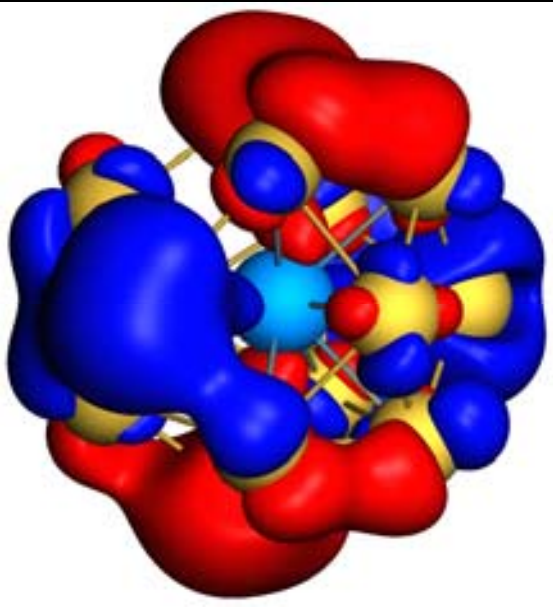<br>HOMOα-5 | 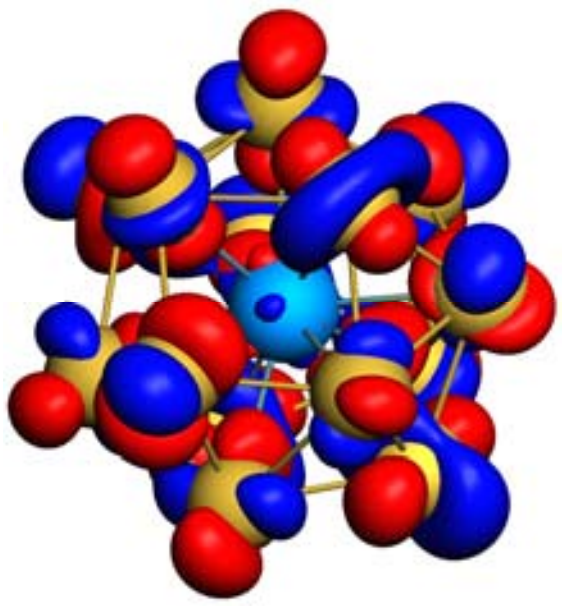<br>HOMOα-30 | 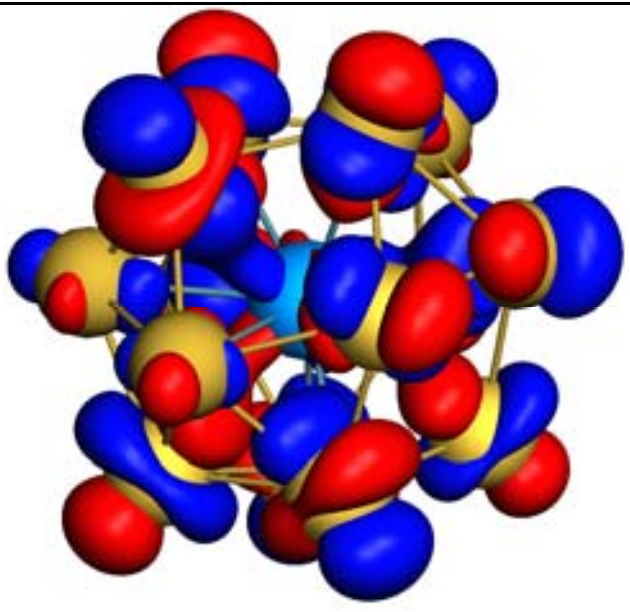<br>HOMOα-34 |
| 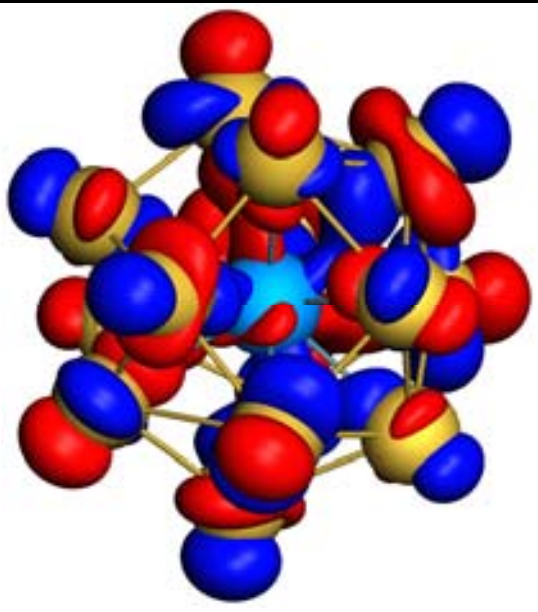<br>HOMOα-35 | 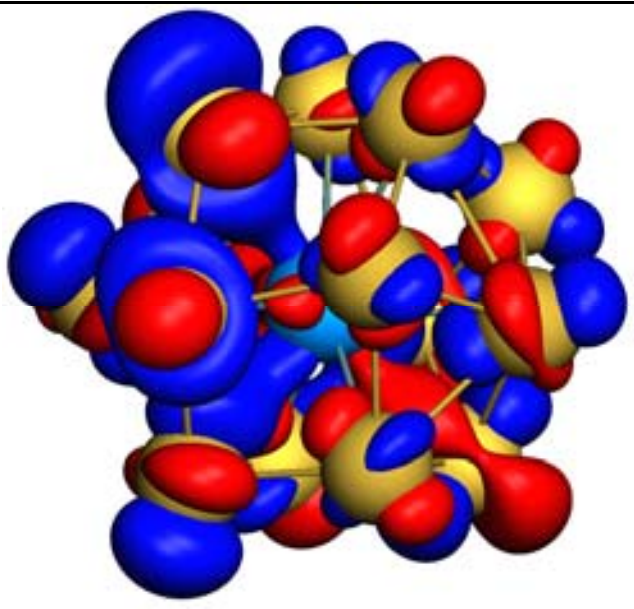<br>HOMOα-37 | 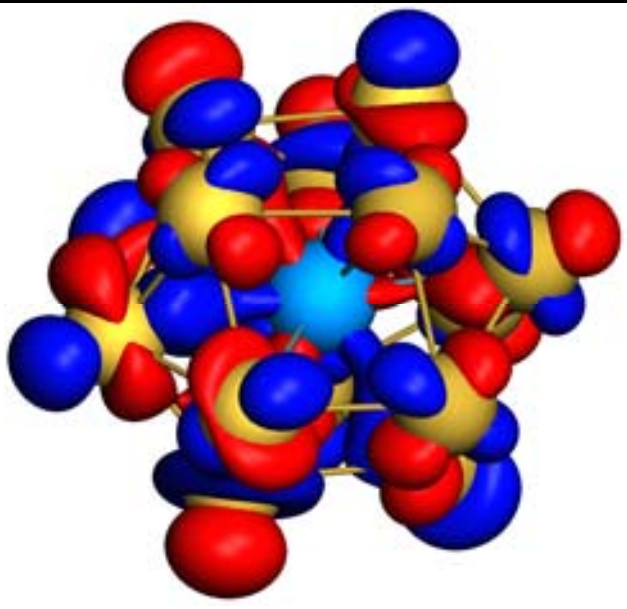<br>HOMOα-39 |
| 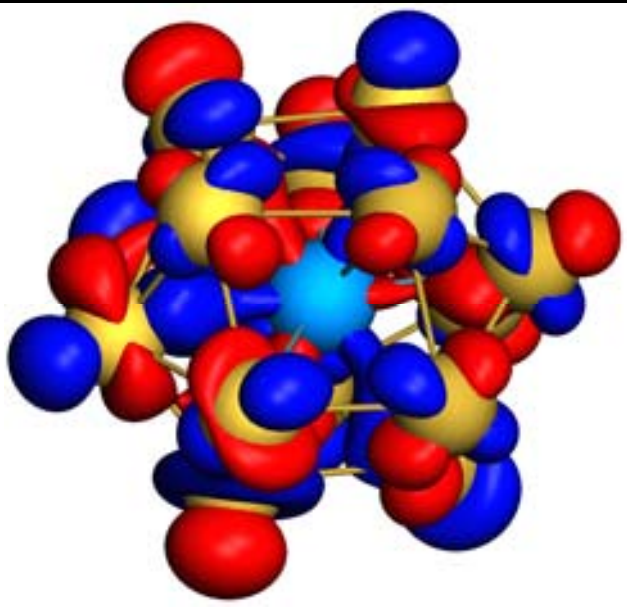<br>HOMOα-41 | 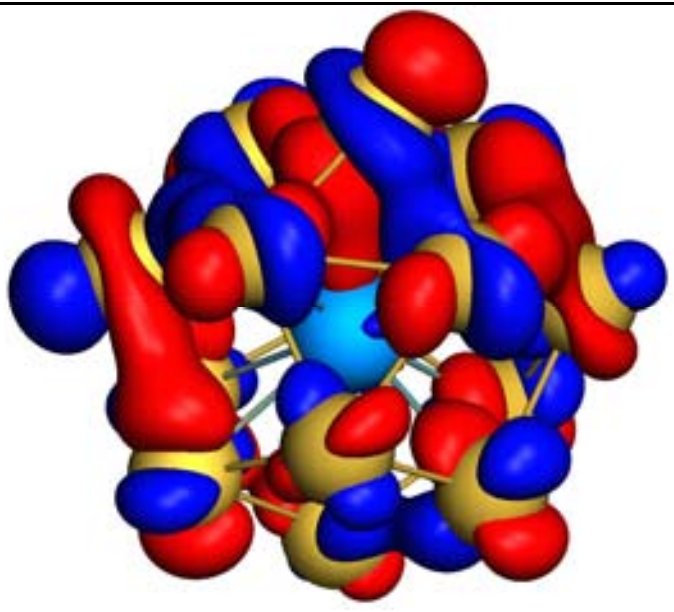<br>HOMOα-45 | 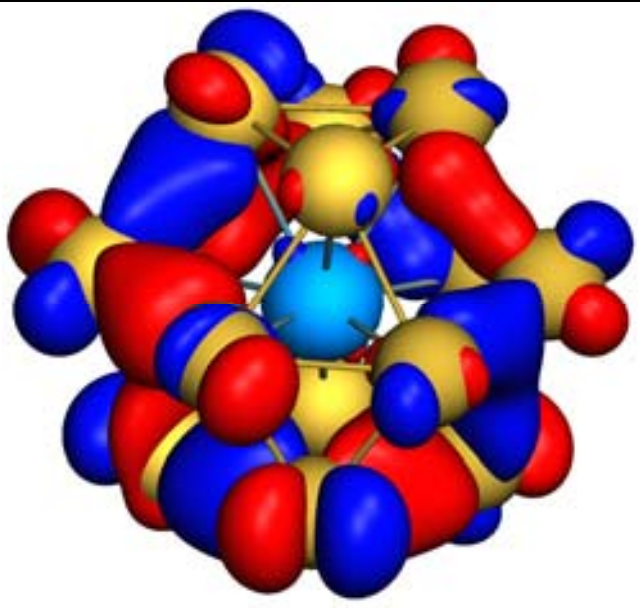<br>HOMOα-58 |
| 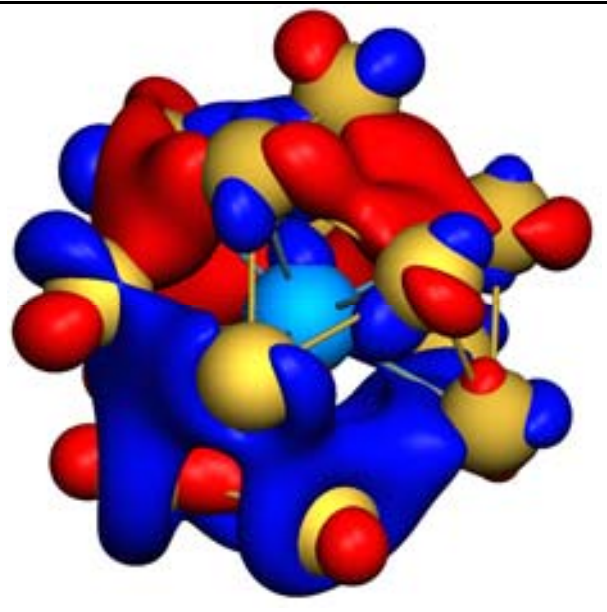 | 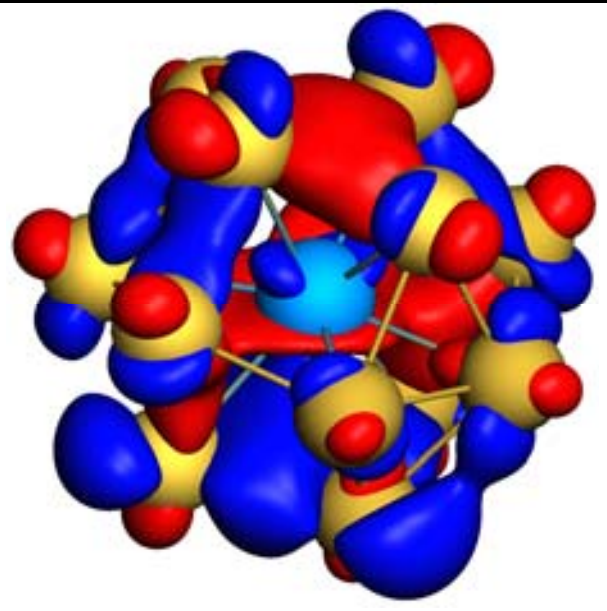 | 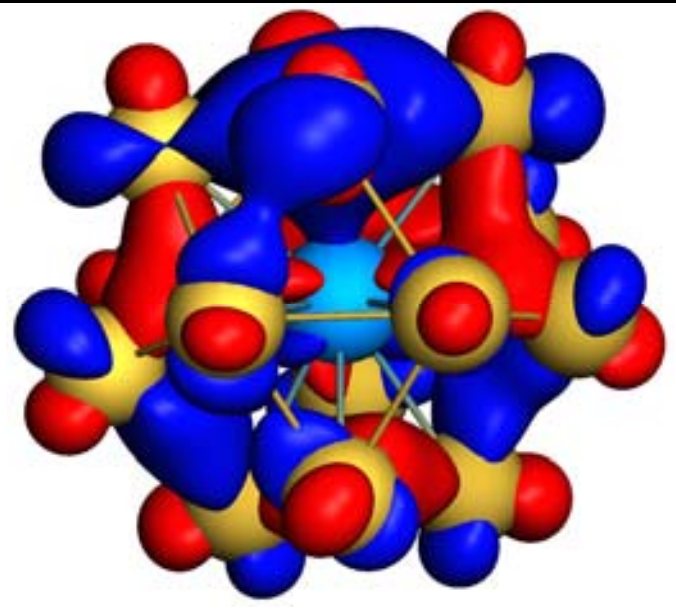 |
| HOMOα-66 | HOMOα-67 | HOMOα-68 |
| 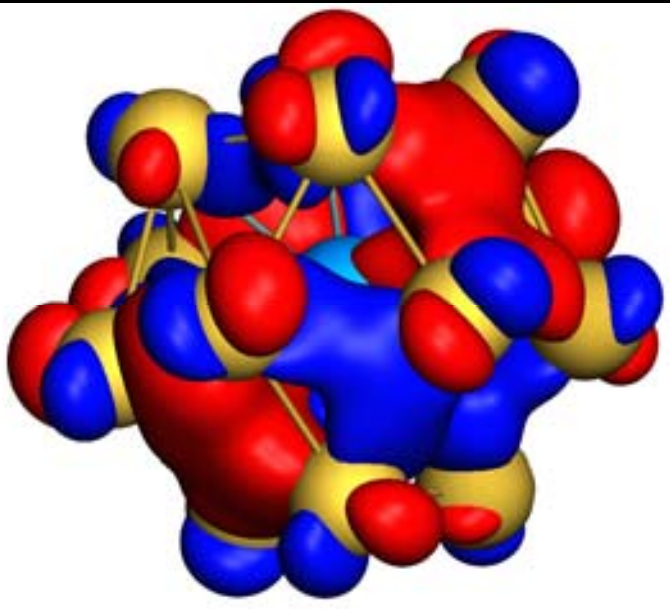 | 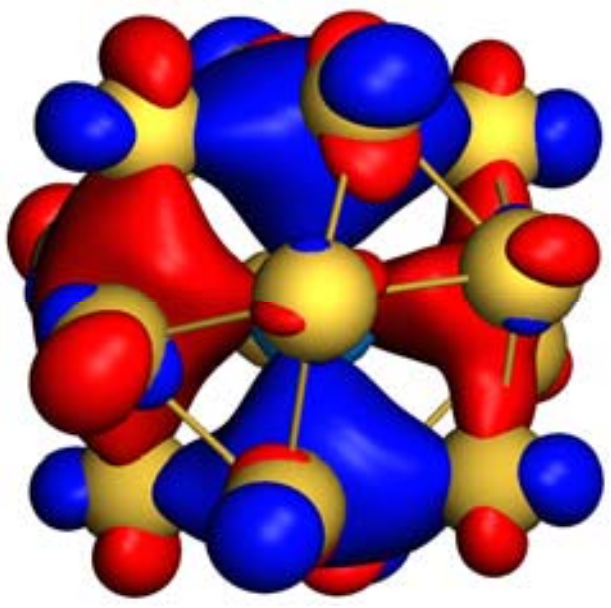 | 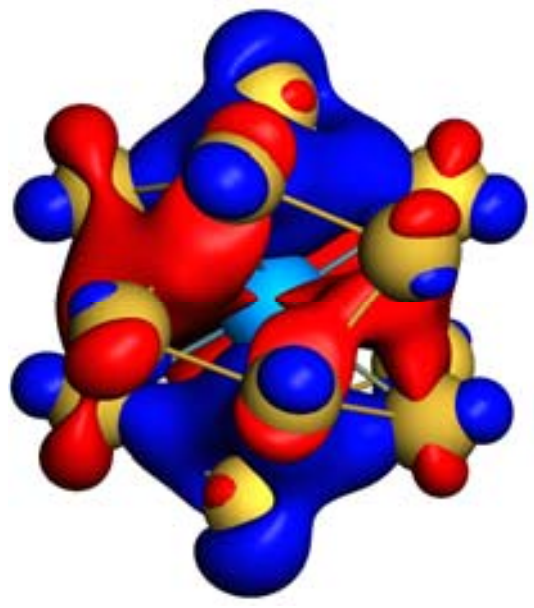 |
| HOMOα-69 | HOMOα-74 | HOMOα-75 |
| 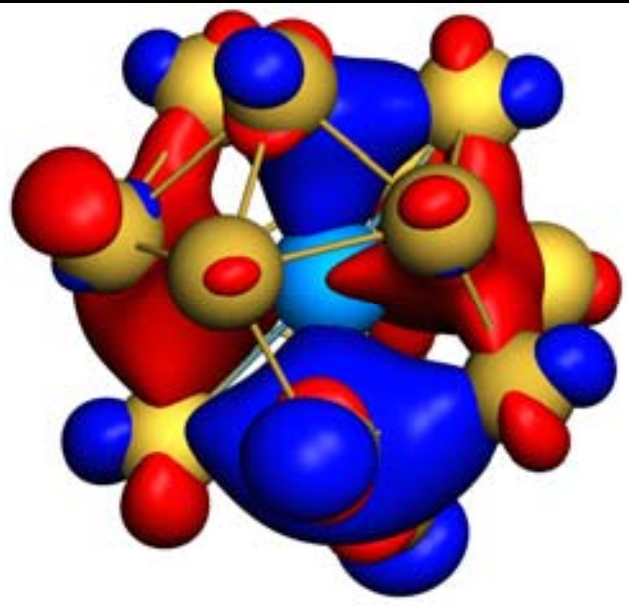 | 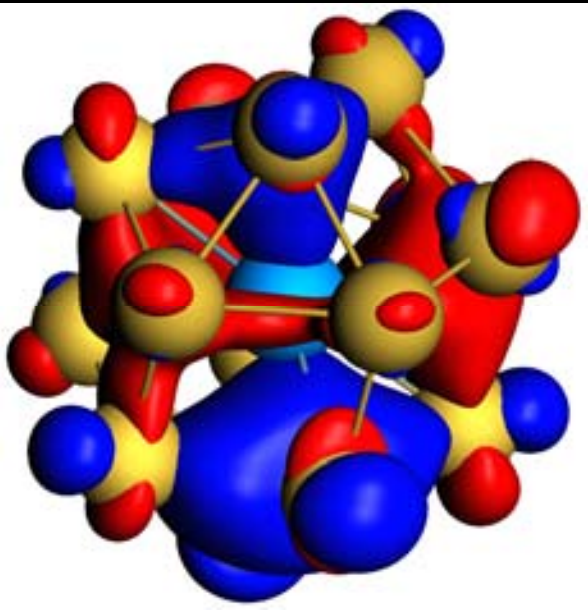 | 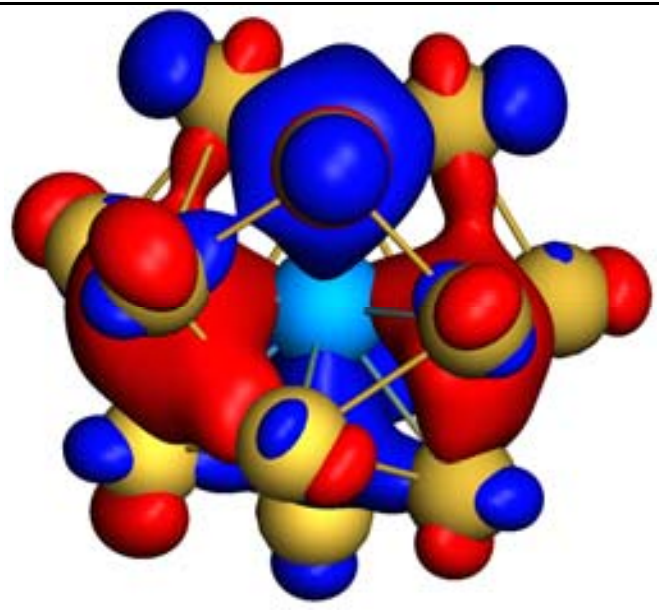 |
| HOMOα-76 | HOMOα-77 | HOMOα-78 |